\documentclass{jetpl}
\usepackage[cp1251]{inputenc}
\usepackage[russian]{babel}
\usepackage{bm}
\usepackage{amsmath}
\usepackage{amssymb}
\usepackage{amscd}
\usepackage[dvips]{graphicx}
\usepackage{epsfig}

\twocolumn

\title{Pair Scattering of Electrons in Edge Channels of Opposite Chiralities
in the Presence of a Disorder Potential}

\rtitle{Pair Scattering of Electrons in
\ldots}

\sodtitle{Pair Scattering of Electrons in Edge Channels of Opposite Chiralities
in the Presence of a Disorder Potential}

\author{M. G. Prokudina, prok@issp.ac.ru, V. S. Khrapai, dick@issp.ac.ru}

\rauthor{M. G. Prokudina, V. S. Khrapai}

\sodauthor{M. G. Prokudina, V. S. Khrapai}

\address{Institute of Solid State Physics, Russian Academy of Sciences, Chernogolovka, Moscow region, 142432 Russia}

\abstract{The nonequilibrium transfer of the energy between electrons of counter-propagating quasi-one-dimensional
systems has been perturbatively calculated for edge channels in a two-dimensional system in the integer quantum Hall effect. The processes involving two electrons that are allowed only in the system with disorder have
been taken into account. Expressions for the cases of Coulomb scattering and transfer of nonequilibrium
phonons have been obtained. The energy transferred per unit time has a quasi-threshold dependence on the
degree of nonequilibrium of the hot channel. According to numerical estimates for electrons in GaAs, Coulomb scattering processes dominate in the energy transfer and the expected effect can be experimentally
observed.}

\begin{document}

\maketitle

The momentum and energy conservation laws in
the one-dimensional case impose significant constraints on electron–electron scattering. For example,
the Coulomb drag effect between single-channel
quantum wires is resonant and maximal when the
interacting systems are identical~\cite{1}. In the regime of
the integer quantum Hall effect, where the diagonal
conductivity in the bulk of the two-dimensional system is exponentially small, chiral quasi-one-dimensional systems in which backscattering is suppressed
(edge channels) can exist~\cite{2}. In the edge channel, all
electrons have the same direction of the momentum
(chirality sign) and drag current without transfer of
particles from one channel to another cannot be
detected at least for a linear dispersion law~\cite{16}. Nevertheless, there is an experimental method for 
measuring the energy transferred in the inelastic scattering of
electrons in different channels~\cite{3,4}.

The pair scattering of electrons in co-propagating edge
channels is forbidden by the momentum conservation
law except for the cases where the dispersion laws are
the same or where disorder providing uncertainty in
the momentum exists in a system~\cite{5}. Energy transfer
between two electrons in the counter-propagating edge
channels in a pure system is strictly forbidden. This
concerns both the Coulomb scattering and transfer of
a nonequilibrium phonon when one electron absorbs a
phonon emitted by another electron. In this case,
three-particle processes that are allowed in the presence of the nonlinear dispersion law should dominate~\cite{6,7}.
However, the pair interaction between counter-propagating edge channels can exist in a non-ideal
system where the role of disorder is significant.

In this work, we study energy transfer between
counter-propagating edge channels one of which has a
nonequilibrium distribution and the second channel is
initially in equilibrium in the presence of smooth disorder using the idea proposed in~\cite{5}. We consider both
the direct unscreened Coulomb interaction and the
interaction through the transfer of nonequilibrium
acoustic phonons. Expressions for the scattering
matrix elements are obtained in perturbation theory.
Numerical estimates are given for edge channels in a
two-dimensional system based on GaAs.

We consider two edge channels with length $L$~\cite{8} with
the linear dispersion law $E(k)=\hbar v_Fk$ ($v_F$ is the drift
velocity at the edge) at distance $d$ from each other.
Channels are separated by a thin barrier impenetrable
for electrons and have opposite chiralities. We assume
that the integer quantum Hall effect for simplicity with
one filled Landau level ($\nu=1$) is implemented in the
bulk of the two-dimensional system. The energy gap
between Landau levels is the largest scale in the problem. Moreover, we neglect the modulation of the 
density of states at the edge and the corresponding edge
features~\cite{9}. This implies that the linear dispersion law
is valid for all energies at which the distribution function
 noticeably differs from the equilibrium distribution function.

As is known, an isolated edge channel in the integer Quantum Hall effect is described in a Fermi liquid model~\cite{10}. A system of two interacting counter-propagating edge channels represents a realization of the Luttinger liquid (LL) and is generally solved by use of a bosonization technique. Energy relaxation in the LL is related to a finite life-time of the bosons and is absent in a clean system~\cite{gutman}. Nevertheless, a finite size of the interaction region gives rise to boson scattering (and energy relaxation) at the boundaries of the LL~\cite{gutman}. We neglect this effect compared to a disorder scattering inside the LL, which is justified given a smallness  of the interaction parameter, not too weak disorder and/or not too small length of the interaction region. Below, the size of the interaction region is assumed to be small compared to the energy relaxation length (boson mean-free path) and a pertubative solution is suggested.

\begin{figure}
\begin{center}
\includegraphics[width=0.6\columnwidth]{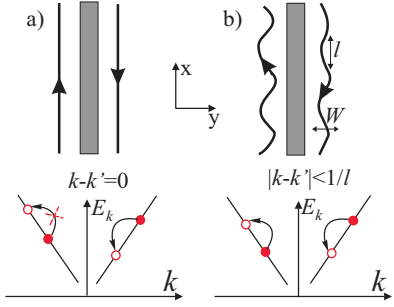}
\end{center}
\caption{{\bf Fig.~1} Schematics of the counter-propagating edge channels separated by an impenetrable barrier and the electron
spectrum for (a) an ideal system (the momentum conservation law is satisfied and pair scattering is forbidden) and
(b) a system with disorder (the momentum conservation
law is violated and pair scattering is allowed).} \label{fig1}
\end{figure}

The momentum and energy conservation laws forbid
pair Coulomb scattering in a pure system (see Fig. 1a).
In the presence of smooth disorder, the equipotential
surface of the edge channel is curved. As a result,
uncertainty appears in the momentum and the inelastic scattering of electrons in neighboring channels
becomes possible. In our model, the deviation of the
edge channel from a straight line is specified by a
random value $\delta d(x)$ with the correlation function
\mbox{$\overline{\delta d(x)\delta
d(x')}=w^2exp\Big(-\frac{(x-x')^2}{2l^2}\Big)$}, where $l$ is the correlation length of smooth disorder (larger than the
magnetic length) and $w$ is the characteristic deviation
(see Fig. 1b). The chosen Gaussian form of the correlation function is not as important as the presence of a
large enough correlation length $l$. In such a system, the
wave-function of quasi-one-dimensional electrons will
depend on both coordinates $x$ and $y$. Since characteristic variations of the momentum are much smaller
than the inverse magnetic length $(eB/(\hbar c))^{1/2}$, this
dependence can be written in the simplified form:

$$\psi_k(x_j,y_j)=\frac{e^{-ikx_j}}{\sqrt{L}}f(y_j-y_0^j(x_j)),$$

Here, subscript $j=1,2$ specifies the channel, $f$ is a narrow function normalized as $\int^{+\infty}_{-\infty}f^2dy_j=1$, and $y_0^j(x_j)=\pm d/2+\delta
d_j(x_j)$, where $\delta d_j(x_j)$  is the random deviation
from the average value $\overline{y_0^j}=\pm d/2$ (the upper and lower
signs correspond to the first and second channels,
respectively) and d is the average distance between the
channels. Owing to this choice of the wave-function,
the matrix element of pair scattering for the
unscreened Coulomb and electron–phonon interactions can be represented in an explicit form.

We denote the initial and final states of two scattered electrons as $|i\rangle=|k,k''\rangle$ and $|f\rangle=|k',k'''\rangle$,
respectively. For the Coulomb interaction in a
medium with the dielectric constant~$\chi$, the matrix element in the first order of perturbation theory is
obtained in the form:

$$V_{if}=\int\psi^*_{k'}(r_1)\psi^*_{k'''}(r_2)\frac{e^2}{\chi|r_1-r_2|}\psi_{k}(r_1)\psi_{k''}(r_2)dr_1dr_2,$$

where $r_j$ is the coordinates of the electron in the $j$th
channel, $j=1,2$. After the substitution of the wave-functions taking into account the choice of $f(y)$, 
the integrand will depend only on $x_1$ and $x_2$, because $\big(y_1-y_2\big)^2=\big(d+\delta d_1(x_1)+\delta
d_2(x_2)\big)^2\approx d^2+2d(\delta d_1+\delta d_2)$. We
change the coordinates $x_1$ and $x_2$ to the relative coordinate $r_x$ and the coordinate of center of mass $R_x$. Owing
to the energy conservation law, integration with
respect to the relative coordinate $r_x$ occurs at zero
momentum. We are interested in scattering with nonzero energy transfer $\varepsilon\neq0$, where the total momentum
is not conserved and the momentum transfer is $\Delta k=k+k''-k'-k'''=2\varepsilon/(\hbar v_F)$. The matrix element has
the form:

$$V_{if}=\frac{e^2d}{\chi L^2}\int\frac{\delta d_1+\delta d_2}{(r_x^2+d^2)^{3/2}}e^{i\Delta
kR_x}dR_xdr_x.$$

where $\delta d_j$ depend on the both variables $R_x$ and $r_x$. It is
easy to obtain an expression for the averaged square of
the absolute value of the matrix element. Taking into
account that \mbox{$\overline{\delta d_i(x)\delta
d_j(x')}=w^2exp\Big(-\frac{(x-x')^2}{2l^2}\Big)$} for
$i=j$ and 0 for $i\neq j$ because the random variables $\delta d_1$ and $\delta d_2$ are not correlated, we obtain:

\begin{equation}
  \overline{|V_{if}|^2}=\frac{8e^4w^2l\sqrt{2\pi}}{\chi^2d^2L^3}e^{-\frac{1}{2}\varepsilon^2/\varepsilon_0^2}\Bigg[\frac{|\varepsilon|}{\varepsilon_d}K_1\Bigg(\frac{|\varepsilon|}{\varepsilon_d}\Bigg)\Bigg]^2,
\label{coulombampl}
\end{equation}

Here, $K_1$ is the modified Bessel function of the second
kind, $\varepsilon_0=\hbar v_F/2l$, and $\varepsilon_d=\hbar v_F/2d$ . At $|\varepsilon|\ll\varepsilon_d$, the
expression in square brackets is close to 1 and the
asymptotic expression $\approx\sqrt{\pi|\varepsilon|/2\varepsilon_d}e^{-|\varepsilon|/\varepsilon_d}$ can be used
for $|\varepsilon|>\varepsilon_d$.

When thermodynamic equilibrium in one channel
is violated, energy can be transferred to a neighboring
channel and the channel can be heated. Let $E,E'$ and
$E'',E'''$ be the initial (final) energy of the electrons in
the first (nonequilibrium) and second channels, respectively. The energy loss of the first electron is $\varepsilon=E-E'$  
(it can formally be negative, corresponding to
an increase in the energy of the first electron). We
denote the distribution function in the channels at the
corresponding energies as $f(k)\equiv  f$ and $f(k')\equiv
f'$. The probability of interaction per unit time is specified by
Fermi’s golden rule $W_{i\rightarrow
f}=\frac{2\pi}{\hbar}\overline{|V_{if}|^2}\delta(\varepsilon+E''-E'''),$,
where $\delta$ is the Dirac delta function. In this case, the
power of the energy transfer through pair scattering is:

\begin{eqnarray}
& &{}P=\frac{2\pi}{\hbar}\frac{L^4}{(2\pi)^4}\iiiint\varepsilon \overline{|V_{if}|^2}\nonumber
f\big(1-f'\big)\times\\&\times&f''\big(1-f'''\big)\delta(\varepsilon+E''-E''')dkdk'dk''dk'''\nonumber
\end{eqnarray}

Taking into account the linear dispersion law, we
transform this integration to integration with respect
to the energies of scattered electrons. In addition, we
denote $E'''-E''=\varepsilon'$; correspondingly, $\delta(\varepsilon+E''-E''')=\delta(\varepsilon-\varepsilon')$. 
Finally, to simplify the expressions, we introduce form factors $F_j(\varepsilon)=\int f_j(E+\varepsilon)\big(1-f_j(E)\big)dE$ for
the emission ($\varepsilon>0$) and absorption ($\varepsilon<0$) of the
energy, where $f_j$ is the distribution function of electrons in the $j$th channel. After the integration of the $\delta$-function, we obtain the final expression:

\begin{eqnarray}
P&=&\frac{2\pi}{\hbar}\frac{L^4}{(2\pi)^4}(\hbar
v_F)^{-4}\times\label{eq:1}\\&\times&\int_{-\infty}^{+\infty}
\varepsilon F_1(\varepsilon)F_2(-\varepsilon)\overline{|V_{if}|^2}d\varepsilon. \nonumber
\end{eqnarray}

We now consider pair scattering via exchange
by nonequilibrium acoustic phonons. In this case, the
scattering probability amplitude should be sought in
the second order of perturbation theory in the electron–phonon interaction:

\begin{equation}
V_{if}=\sum_{q_x,q_y,q_z} \frac{\langle f|V_{e-ph}|I\rangle\langle I|V_{e-ph}|i\rangle}{E_i-E_I+i0},
\label{secondorder}
\end{equation}

where the intermediate state $|I\rangle=|k',k'',{\bf q}\rangle$  corresponds to the relaxation of the first electron with the
emission of a real or virtual phonon and summation is
performed over the three-dimensional momenta of
phonons and their polarizations. We assume that the
phonon system is closely connected to a low-temperature thermostat so that the steady-state phonon
occupation numbers at energies of interest are much
smaller than 1. Bearing in mind a GaAs crystal, we
take into account the electron–phonon interaction
through piezoelectric coupling, which dominates in
the long-wavelength limit. In the case of the emission of a phonon by an electron in the first channel and its
absorption in the second channel, we obtain:

\begin{eqnarray}
\langle f|V_{e-ph}|I\rangle\langle I|V_{e-ph}|i\rangle=\frac{\alpha\Omega}{L^2Vq}e^{iq_yd}\times\nonumber\\
\times\int^L_{0}e^{i\Delta k_1x}e^{iq_y\delta d_1(x)}dx\int^L_{0}e^{i\Delta k_2x}e^{-iq_y\delta d_2(x)}dx,
\label{nominator}
\end{eqnarray}

where $\Omega\equiv\hbar
e^2e_{14}^2/(2\chi^2\rho s)$; $\rho$ and $V$ are the density and
volume of the crystal, respectively; and $e_{14}$ is the piezoelectric coefficient in a crystal with a zinc blende
structure~\cite{11}. The speed of sound $s$ and coefficient $\alpha$
depend on the direction of the momentum of a
phonon and its polarization~\cite{12}. Below, estimates will
be given for edge channels along the [110] easy cleavage direction in a two-dimensional gas in GaAs. In this
case, the main contribution comes from transverse
acoustic phonons with $\alpha=0.5$ and $s=3\times10^5$~cm/s.
The non-conserved longitudinal momentum at emission ($j = 1$) and absorption ($j = 2$) of a phonon is
conveniently represented in the form $\Delta k_j=\varepsilon/\hbar v_F+(-1)^jq_x$, where the energy conservation law is taken
into account. In our model, the effective width of the
edge channel is $w$; consequently, the $y$ component of
the phonon momentum is bounded as $|q_y|\lesssim1/w$. For
this reason, the approximate expression $e^{-(-1)^jiq_y\delta d_j(x)}\approx1-(-1)^jiq_y\delta d_j(x)$, $j=1,2$ can be used in Eq.(\ref{nominator}).
Then, Eq.(\ref{secondorder}) for the matrix element has the form:

\begin{eqnarray}
V_{if}=\frac{\alpha\Omega}{8\pi^3L^2}\int\frac{e^{iq_yd}dq_xdq_ydq_z}{q(\varepsilon-\varepsilon_q+i0)}\Bigg[q_y^2I_1I_2+\label{amplitude}\\+i2\pi q_y\delta(q_x+\varepsilon/\hbar v_F)I_1-i2\pi q_y\delta(q_x-\varepsilon/\hbar v_F)I_2\Bigg],\nonumber
\end{eqnarray}

where $\delta(q_x\pm\varepsilon/\hbar v_F)$ is the delta function, $\varepsilon_q=\hbar sq$ is
the energy of the phonon, and $I_j=\int e^{i\Delta k_jx}\delta d_j(x)dx$. The
last two terms in square brackets correspond to processes with single violation of the momentum conservation law at the emission or absorption of a phonon.
The first term concerns processes where the
momentum is not conserved at both the emission
and absorption of a phonon. The integrals ${\rm Int}_n(\varepsilon)=\int (q_y)^ne^{iq_yd}dq_ydq_z/q/(\varepsilon-\varepsilon_q+i0)$ for $n=1,2$ should be
calculated in Eq.~(\ref{amplitude}). It is convenient to change the
variables $dq_ydq_z=q_\perp dq_\perp d\varphi$, where $q_\perp=\sqrt{q_y^2+q_z^2}$  is the
phonon momentum component perpendicular to the
x axis and $\varphi$ is the angle between $\bf q_\perp$ and the z axis.
Then, at $s\ll v_F$, $q_\perp\approx q$ because acoustic phonons
under these conditions are emitted primarily along the normal to the edge channel. Changing $q\rightarrow\varepsilon_q/(\hbar s)$,
we finally obtain:

\begin{equation}
{\rm Int}_n(\varepsilon)\approx(\hbar s)^{-n-1}\int\sin^n{\varphi}d\varphi \int\frac{\varepsilon_q^ne^{i\varepsilon_q\tau\sin\varphi}d\varepsilon_q}{\varepsilon-\varepsilon_q+i0},
\label{integrals}
\end{equation}

where $\tau=d/s$ is the characteristic time of flight of a
phonon between the channels. In Eq.~(\ref{integrals}), the integral
with respect to the energy of the phonon over the range
$0<\varepsilon_q<\hbar s(w|\sin\varphi|)^{-1}$ is taken first and the integral with respect to $\varphi$ is then calculated. First, we
present an answer in the limit $\varepsilon\gg\hbar/\tau$; for the characteristic values $\varepsilon\sim\varepsilon_0$, this limit is equivalent to a long
distance between the edge channels $d\gg ls/v_F$. As will be
shown below, the last condition is certainly satisfied
when the electron.phonon mechanism can compete
with the Coulomb mechanism. In this limit, the main
contribution to the integral comes from phonons in a
narrow energy band $|\varepsilon_q-\varepsilon|\lesssim \hbar/\tau$. As a result,

\begin{equation}
{\rm Int}_n(\varepsilon)\approx\frac{2\pi^2\varepsilon^n}{(i\hbar s)^{n+1}}\frac{d^n}{da^n}\Bigg(J_0(a)+iH_0(a)\Bigg)\Bigg|_{a=\varepsilon\tau/\hbar},
\label{dinfty}
\end{equation}

where $J_0$ and $H_0$ are the Bessel function of the first kind
and the Struve function of the zeroth order, respectively, and the derivative of the quantity in square
brackets is taken at the indicated argument. In the
limit $a\rightarrow\infty$, the asymptotic expression $J_0(a)+iH_0(a)\approx\sqrt{2/\pi a}e^{i(a-\pi/4)}$ can be used.

In the opposite limit of small distances, $\varepsilon\ll\hbar/\tau$, all
phonons contribute to the integral. As a result, the
approximate answer has the form

\begin{eqnarray}
{\rm Int}_1\approx \frac{2i\pi(\cos(d/w)-1)}{\hbar sd}\nonumber\\
{\rm Int}_2\approx \frac{-2\pi\sin(d/w)}{\hbar swd}\nonumber
\end{eqnarray}

It only remains to calculate the integral with
respect to $q_x$. For two last terms in Eq.~(\ref{amplitude}), the integral
is easily calculated owing to the $\delta$-function. However,
for the first term, the square of the absolute value of the
matrix element should first be taken and the resulting
double integral with respect to $q_x$ and $q'_x$ should then
be calculated. Finally, averaging over the random distributions $\delta d_j$ should be performed, similar to the case
of Coulomb scattering. As a result,

$$\overline{|V_{if}|^2}=V_1+V_2, {\rm where}  $$

\begin{eqnarray}
V_1=2^{7/2}\left(\frac{\alpha\Omega}{8\pi^3L^2}\right)^2|{\rm Int}_1|^2\pi^{5/2}w^2Lle^{-\varepsilon^2/2\varepsilon_0^2}
\nonumber \\ \label{amplsquared} \\
V_2=4\left(\frac{\alpha\Omega}{8\pi^3L^2}\right)^2|{\rm Int}_2|^2\pi^{5/2}w^4Lle^{-\varepsilon^2/4\varepsilon_0^2}
\nonumber
\end{eqnarray}

Here, subscripts 1 and 2 correspond to the single
and double violations of the momentum conservation
law, respectively. It is noteworthy that Eqs.~(\ref{amplsquared}) are also
valid for the transfer of a phonon from the second
channel to the first one ($\varepsilon<0$) if change $\varepsilon\rightarrow|\varepsilon|$ is
made in Eq.~(\ref{dinfty}). The final expression for the power
transferred between the edge channels through the
electron–phonon interaction can be easily obtained
from Eq.~(\ref{eq:1}).

It is interesting to compare the efficiencies of two
interaction mechanisms. At low energies $|\varepsilon|<\varepsilon_d,\varepsilon_0$ and/or small distances between the channels, $d\lesssim l$,
the matrix element for exchange by phonons is negligible compared to the Coulomb matrix element.
However, the phonon contribution becomes noticeable at $d>>l$ and relatively high energies $|\varepsilon|>\varepsilon_d$. Indeed,
the energy cutoff in Coulomb matrix element~(\ref{coulombampl})
occurs at $|\varepsilon|\sim\varepsilon_d\ll\varepsilon_0$, i.e., much earlier than in the
electron-phonon matrix element given by Eq.~(\ref{amplsquared}). As
a result, in the case of the strongly nonequilibrium first
channel and the weakly nonequilibrium second channel (see below), the power transferred through 
Coulomb scattering decreases rapidly with the distance
between the channels, $P_C\propto1/d^5$. At the same time, the
asymptotic behavior of the phonon contribution is
$P_{ph}\propto1/d$. The ratio of the powers is given by the
expression

\begin{eqnarray}
\frac{P_{ph}}{P_C}=\frac{e_{14}^4v_F}{6\pi^2\rho^2s^5}\Bigg(1+8\sqrt{2}\frac{v_F^2}{s^2}\frac{w^2}{l^2}\Bigg)\cdot\frac{d^4}{l^4}
\nonumber
\end{eqnarray}

In a clean system, one should  expect that $w/l\ll1$
owing to the high electric field at the edge. If $w/l=0.1$ is
taken for estimates, the phonon contribution becomes
comparable to the Coulomb one when the
distance between the channels satisfies the condition
$d/l\approx10$.

Below, we consider in more detail the case of pair Coulomb scattering in the limit of small distances
between the channels, $d\ll l$, where the integrand in square brackets in~(\ref{coulombampl}) can be taken as 1. Solution~(\ref{eq:1})
makes it possible to obtain a numerical result for the power transfer for arbitrary distribution functions of
electrons in the first and second channels. In the limiting case of the strongly nonequilibrium first channel
and the weakly nonequilibrium second channel, the result can be represented in an explicit form. The integral with respect to the energy in Eq.~(\ref{eq:1}) has cutoff at $|\varepsilon|\sim\varepsilon_0$. If nonequilibrium in the first channel has a
noticeably larger energy scale, the energy dependence can be neglected, i.e., $F_1(\varepsilon)\approx F_1(0)$. For example, for
the quasi-equilibrium distribution with the effective temperature $F_1(0)=k_BT_1$ (where $k_B$ is the Boltzmann constant). For the two-step distribution function created by a quantum contact with the applied voltage $V$
and transmission coefficient $Tr$: $F_1(0)=Tr(1-Tr)e|V|$. The condition of strong nonequilibrium corresponds
to the inequality $k_BT_1,e|V|\gg\varepsilon_0$. At the same time,
when the temperature in the second channel is low,
$k_BT_2\ll\varepsilon_0$, $F_2(-\varepsilon)\approx\varepsilon\theta(\varepsilon)$, where $\theta$ is the Heaviside
step function (i.e., energy in the second channel can only be absorbed). In this limit, solution~(\ref{eq:1}) has the form

\begin{equation}
P_C=\frac{e^4w^2L}{8\pi^2\hbar^2v_F\chi^2d^2l^2}F_1(0)
\nonumber
\end{equation}

Therefore, in the strongly nonequilibrium case, the power transferred via pair scattering processes
depends linearly on the drive voltage or on the effective temperature of the hot channel~\cite{14} (this statement is valid for both interaction mechanisms and an arbitrary distance between the channels). In the other limiting case of weak nonequilibrium in both channels ($k_BT_2\ll k_BT_1,e|V|\ll\varepsilon_0$) $P_C\propto (T_1)^4, V^4$. Thus, the dependence of the power on the overdrive voltage or the effective temperature of the hot channel
has a quasi-threshold behavior with the threshold $k_BT_1,eV\sim\varepsilon_0$.

Finally, we estimate the scale of the effect in a relatively clean system, assuming that the initial temperature
$T_2^{in}$ of electrons in the second channel before interaction is low. When the distance between the channels is $d=300$nm, $v_F=10^7$cm/s, disorder parameters are $l=1\mu$m and $w=50$nm, and nonequilibrium in the first
channel is created by a quantum point contact with $Tr=1/2$, the effective temperature in the second channel at
the exit from the interaction region is~\cite{15}:

$$T_2^{out}=\sqrt{\frac{12\hbar P_C}{\pi k_B^2}+(T_2^{in})^2}.$$

For the interaction length $L=3\mu$m, $eV=0.5$meV$\gg\varepsilon_0=0.05$
meV, and $T_2^{in}=50$mK, we obtain $T_2^{out}\approx130$mK. Such a temperature change can be measured
in practice and was possibly observed in~\cite{heiblum}.

To summarize, pair scattering of electrons of one-dimensional Fermi liquids of the opposite chiralities
has been calculated. The energy is transferred between
systems owing to the violation of the momentum conservation law in the presence of a smooth disorder
potential. Expressions for the power transfer in the case of the Coulomb and electron–phonon scattering
mechanisms have been obtained for the edge channels in a two-dimensional GaAs system. At small distances
between the channels, Coulomb scattering dominates and the maximum energy quantum transferred in
interaction is determined by the disorder correlation length similar to~\cite{5}. As a result, the power transfer has
a quasi-threshold dependence on the degree of nonequilibrium (temperature or drive voltage) in the
hot channel.

We are grateful to I.S. Burmistrov, I.V. Gornyi,
D.V. Shovkun, and A.A. Shashkin for stimulating discussions and to D.G. Polyakov for valuable remarks.
This work was supported by the Russian Foundation
for Basic Research, the Russian Academy of Sciences, the Ministry of Education and Science of the
Russian Federation, and the Council of the President of the Russian Federation for Support of Young
Scientists and Leading Scientific Schools (project no. MK-3102.2011.2).


\begin{thebibliography}{20}
\bibitem{8} All formulas are also valid for infinitely long edge channels
interacting in the finite segment with the length $L$.
\bibitem{14} A linear dependence of the power transfer for three-particle
scattering processes on the overdrive voltage is possible only at
$Tr\ll1$. In other cases, the dependence will be at least quadratic.
\end{thebibliography}
\end{document}